\documentclass[aps,prl,10pt,twocolumn,showpacs,superscriptaddress,footnoteinbib]{revtex4}
\usepackage{amsmath}
\usepackage{latexsym}
\usepackage{amssymb}
\usepackage{color}
\usepackage{graphicx}
\usepackage{graphics,epstopdf}
\usepackage{soul}
\usepackage[breaklinks]{hyperref}

\begin{document}

\title{Non-Local Advantage of Quantum Coherence}
\author{Debasis Mondal}
\thanks{debamondal@hri.res.in} 
\affiliation{Quantum Information and Computation Group, Harish-Chandra Research Institute, Chhatnag Road, Jhunsi, Allahabad, India}
\affiliation{Homi Bhabha National Institute,
Anushaktinagar, \\Training School Complex, Mumbai-400085, India.}

\author{Tanumoy Pramanik}
\thanks{Tanumoy.Pramanik@telecom-paristech.fr}
\affiliation{LTCI, T\'{e}l\'{e}com ParisTech, 23 avenue dItalie, 75214 Paris CEDEX 13, France.}

\author{Arun Kumar Pati}
\thanks{akpati@hri.res.in}
\affiliation{Quantum Information and Computation Group, Harish-Chandra Research Institute, Chhatnag Road, Jhunsi, Allahabad, India}
\affiliation{Homi Bhabha National Institute,
Anushaktinagar, \\Training School Complex, Mumbai-400085, India.}
\date{\today}

\begin{abstract}
 
A bipartite state is said to be steerable if and only if it does not have a single system description, i.e., the bipartite state cannot be explained by a local hidden state model. Several steering inequalities have been derived using different local uncertainty relations to verify the ability to control the state of one subsystem by the other party. Here, we derive complementarity relations between coherences measured on mutually unbiased bases using various coherence measures such as the $l_1$-norm, relative entropy and skew information. Using these relations, we derive conditions under which non-local advantage of quantum coherence can be achieved and the state is steerable. We show that not all steerable states can achieve such advantage. 
\end{abstract}


\maketitle

Steering is a kind of non-local correlation introduced by Schr\"{o}dinger~\cite{Schrodinger} to reinterpret the EPR-paradox~\cite{EPR}. According to Schr\"{o}dinger, the presence of entanglement between two subsystems in a bipartite state enables one to control the state of one subsystem by its entangled counter part. Wiseman {\it et al}.~\cite{Jones07} formulated the operational and mathematical definition of quantum steering and showed that steering lies between quantum entanglement and Bell non-locality on the basis of their strength~\cite{Bell}. The notion of the steerability of quantum states is also intimately connected \cite{brezger} to the idea of remote state preparation \cite{rsp,apati}.

As introduced in Ref.~\cite{Jones07}, let us consider a hypothetical game to explain the steerability of quantum states. Suppose, Alice prepares two quantum systems, say, $A$ and $B$ in an entangled state $\rho_{AB}$ and sends the system $B$ to Bob. Bob does not trust Alice but agrees with the fact that the system $B$ is quantum. Therefore, Alice's task is to convince Bob that the prepared state is indeed entangled and they share non-local correlation. On the other hand,  Bob thinks that Alice may cheat by preparing the system $B$ in a single quantum system, on the basis of possible strategies~\cite{Saunders, FUR_St}. Bob agrees with Alice that the prepared state is entangled and they share non-local correlation if and only if the state of Bob cannot be written by local hidden state model (LHS)~\cite{Jones07}
\begin{eqnarray}
\rho_{A}^{a} = \sum_{\lambda} \mathcal{P}(\lambda)\,\mathcal{P}(a|A,\lambda)\,\rho_{B}^Q(\lambda),
\label{LHS}
\end{eqnarray}
where $\{\mathcal{P}(\lambda),\rho_{B}^{Q}\}$ is an ensemble of LHS prepared by Alice and  $\mathcal{P}(a|A,\lambda)$ is Alice's stochastic map to convince Bob.
Here, we consider $\lambda$ to be a hidden variable with the constraint $\sum_{\lambda}\mathcal{P}(\lambda)=1$ and $\rho_{B}^{Q}(\lambda)$ is a quantum state received by Bob. 
The joint probability distribution on such states, $P(a_{\mathcal{A}_i},b_{\mathcal{B}_i})$ of obtaining outcome $a$ for the measurement of observables chosen from the set $\{\mathcal{A}_i\}$ by Alice and outcome $b$ for the measurement of observables chosen from the set $\{\mathcal{B}_i\}$ by Bob can be written as
\begin{eqnarray}
P(a_{\mathcal{A}_i},b_{\mathcal{B}_i}) = \sum_{\lambda} P(\lambda)\, P(a_{\mathcal{A}_i}|\lambda)\, P_Q(b_i|\lambda),
\label{LHS_JP}
\end{eqnarray}
where $P_Q(b_i|\lambda)$ is the quantum probability of the measurement outcome $b_{i}$ due to the measurement of $\mathcal{B}_{i}$.

Several steering conditions have been derived on the basis of Eq.~(\ref{LHS_JP}) and the existence of single system description of a part of the bi-partite systems~\cite{Saunders, FUR_St, Walborn}. It has also been quantified for two-qubit systems \cite{EPR_quant}.
In the last few years, several experiments have been performed to demonstrate the steering effect with the increasing measurement settings~\cite{Saunders} and with loophole free arrangements~\cite{Exp_St_1}. For continuous variable systems, the steerability has also been quantified \cite{gerardo}. 


Recently, quantum coherence has been established as an important notion, specially in the areas of quantum information theory, quantum biology \cite{huelga1,huelga2,rebentrost,lloyds,abbott} and quantum thermodynamics \citep{guzic,gour,rudolph1,rudolph2,gardas}. 
In quantum information theory, it is expected that it can be used as a resource  \cite{l1_norm,girolami,uttam}. This has been the main motivation for recent studies to quantify and develop a number of measures of quantum coherence \cite{girolami,aberg06,l1_norm,singh}. Most importantly, operational interpretations of resource theory of quantum coherence have also been put forward \cite{winter,singh1}. An intriguing connection between quantum coherence and quantum speed limit has been established \cite{deba, spekken}. However, much work needs to be done to really understand how to control and manipulate coherence so as to use it properly as a resource, particularly, in multipartite scenario. 

In this rapid communication, we study the effects of non-locality on quantum coherence in bi-partite scenario. We derive a set of inequalities for various quantum coherence measures. Violation of any one of these inequalities by the conditional states of a part of the system implies that it can achieve non-local advantage (the advantage, which cannot be achieved by a single system and LOCC) of quantum coherence. Moreover, these inequalities can also be considered as sufficient steering criteria. Intuitively, for quantum systems, it may seem that all steerable states can achieve the non-local advantage on quantum coherence. But here we show that for mixed states, steerability captured by different steering criteria~\cite{Saunders, FUR_St, Walborn} based on uncertainty relations are drastically different from the steerability captured by coherence. In other words, we show that there are steerable states, which cannot achieve the non-local advantage of quantum coherence.  

One should note that we do not aim to derive a stronger steering criteria but aim to establish a connection between the steerability and the quantum coherence. This eventually leads us to show the effects of quantum steering on the speed of quantum evolutions (see supplemental material \cite{supp}).

\begin{figure}
\includegraphics[scale=.5]{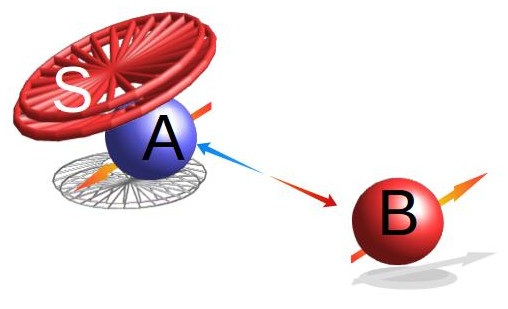}
\label{fig:1}
\caption{Coherence of Bob's particle is being steered beyond what could have been achieved by a single system, only by local projective measurements on Alice's particle and classical communications(LOCC).}
\end{figure} 
To quantify coherence, we consider the $l_1$-norm and the relative entropy of coherence as a measure of quantum coherence~\cite{l1_norm}. We also use the skew information~\cite{Skew}, which is an observable measure of quantum coherence \cite{girolami} and also known as a measure of asymmetry \cite{gour1,gour2,ahmadi,marvian1}. The $l_1$-norm of coherence of a state $\rho$ is defined as $C^{l_1}(\rho) = \sum_{\substack{i,j\\i\neq j}} |\rho_{i,j}|.$ Now, if a qubit is prepared in either spin up or spin down state along z-direction then the qubit  is incoherent, when we calculate the coherence in $z$-basis (i.e., $C_z^{l_1}=0$) and is fully coherent in x- and y-basis, i.e., 
$C_{x\,(y)}^{l_1}=1$. The $l_1$-norm of coherence of a general single qubit $\rho=\frac{1}{2}(I+\vec{n}.\vec{\sigma})$ (where  $|\vec{n}| \leq 1$ and $\vec{\sigma}\equiv(\sigma_{x},\sigma_{y},\sigma_{z})$ are the Pauli matrices) in the basis of Pauli matrix $\sigma_i$ is given by
\begin{eqnarray}
C^{l_1}_i(\rho) = \sqrt{n_j^2+n_k^2},
\end{eqnarray}
where $k\neq i\neq j$ and $i, j, k\in\{x, y, z\}$.

Therefore, one may ask, what is the upper bound of $\mathcal{C}^{l_1}=C_x^{l_1}(\rho)+C_y^{l_1}(\rho)+ C_z^{l_1}(\rho)$ for any general qubit state $\rho$. 
Using $C^{l_1}_x C^{l_1}_y + C^{l_1}_x C^{l_1}_z+ C^{l_1}_y C^{l_1}_z \leq C_x^2 + C_y^2 +C_z^2\,\leq\, 2$ (see \cite{supp}), we find that the above quantity 
is upper bounded by
\begin{eqnarray}
\sum_{i=x,y,z}C_i^{l_1}(\rho)\leq \sqrt{6},
\label{C_Uncer_l1}
\end{eqnarray} 
where the equality sign holds for a pure state, which is an equal superposition of all the mutually orthonormal states 
spanning the state space, i.e.,
\begin{eqnarray}
\rho^{\mathcal{C}}_{\max}=\frac{1}{2}\,\Big[I \,+ \,\frac{1}{\sqrt{3}} (\sigma_x\,+\,\sigma_y\,+\,\sigma_z) \Big],
\label{Max_C_State}
\end{eqnarray}
where $I$ is $2\times 2$ identity matrix. Hence, in the single system description, the quantity $\mathcal{C}^{l_1}$ 
cannot be larger than $\sqrt{6}$ and the corresponding inequality~(\ref{C_Uncer_l1}) can be thought as a coherence 
complementarity relation. 

Another measure of coherence called the relative entropy of coherence is defined as~\cite{l1_norm}
$C^{E}(\rho)= S(\rho_D)\,-\,S(\rho),$
where $S(\rho)$ is the von-Neumann entropy of the state $\rho$ and $\rho_D$ is the diagonal matrix formed by the diagonal elements of $\rho$ in a fixed basis, i.e., $\rho_D$ is completely decohered state of $\rho$. This quantity has also been considered as `wavelike information' in Ref. \cite{angelo}, which satisfies a duality relation. In this case, the sum of coherences of single qubit system in the three mutually unbiased bases for qubit systems is bounded by
\begin{eqnarray}
\sum_{i=x,y,z}C_i^{E}(\rho)  &=& \sum_{i=x,y,z}\mathcal{H}\left(\frac{1+n_i}{2}\right) - 3\mathcal{H}\left(\frac{1+|\vec{n}|}{2}\right)\nonumber\\
& \leq& C^{m}_{2},
\label{C_Uncer_RE}
\end{eqnarray}
where $\mathcal{H}(x)=-x\log_2(x)-(1-x)\log_2(1-x)$ and $|\vec{n}|=\sqrt{n_{x}^2+n_{y}^2+n_{z}^2}$. Using the symmetry, one can easily show that the maximum occurs at $n_x=n_y=n_z=1/\sqrt{3}$ (i.e., for maximally coherent state given by Eq.~(\ref{Max_C_State})) and  $C^{m}_{2}\approx 2.23$.

Recently, the skew-information \cite{Skew} has also been considered as an observable measure of coherence of a state 
\cite{girolami}. The coherence of a state $\rho$, captured by an observable $\mathcal{B}$, i.e., the coherence 
of the state in the basis of eigenvectors of the spin observable $\sigma_i$ is given by
\begin{eqnarray}
C^S_{i} = -\frac{1}{2} Tr[\sqrt{\rho},\sigma_i]^2 = \frac{\left(n_j^2+n_k^2\right)\left(1-\sqrt{1-|\vec{n}|^2}\right)}{|\vec{n}|^2},
\end{eqnarray}
which is a measure of quantum part of the uncertainty for the measurement of the observable $\sigma_i$ and hence it does not increase under classical mixing of states~\cite{Skew}. The sum of the coherences measured by skew information in the bases of $\sigma_x$, $\sigma_y$ and $\sigma_z$ is upper bounded by
\begin{eqnarray}\label{un_skew}
\sum_{i=x,y,z}C_{i}^{S}(\rho) = 2\left(1-\sqrt{1-|\vec{n}|^2}\right) \leq 2,
\label{C_Uncer_S}
\end{eqnarray}
where the maximum occurs for maximally coherent state $\rho^{\mathcal{C}}_{\max}$ given by Eq.~(\ref{Max_C_State}).
The inequalities~(\ref{C_Uncer_l1}), (\ref{C_Uncer_RE}) and (\ref{C_Uncer_S}) are complementarity relations for coherences of a state measured in the mutually unbiased bases. 

Let us now describe our steering protocol, which we use to observe the effects of steering of the coherence of a part of a bi-partite system. We consider a general two-qubit state of the form of
\begin{eqnarray}
\eta_{AB}= && \frac{1}{4} (I^A\otimes I^B+\vec{ r} \cdot {\sigma}^{A}\otimes I^B+  I^{A}\otimes \vec{s}\cdot \vec{\sigma}^{B} \nonumber \\
&&+\sum_{i,j=x,y,z} t_{ij} \sigma_{i}^{A}\otimes \sigma_{j}^{B}),
\label{G_2qbit}
\end{eqnarray}
where $\vec{r}\equiv(r_{x},r_{y},r_{z})$, $\vec{s}\equiv(s_{x},s_{y},s_{z})$, 
with $|r|\leq 1$, $|s|\leq 1$ and $(t_{ij})$ is the correlation matrix.
Alice may, in principle, perform measurements in arbitrarily chosen bases. For simplicity, we derive the coherence steerability criteria for three measurement settings in the eigenbases of $\{\sigma_{x}, \sigma_{y}, \sigma_{z}\}$. 
 When Alice declares that she performs measurement on the eigenbasis of $\sigma_z$ and obtains outcome $a\in\{0,1\}$ with probability $p(\eta_{B|\Pi_{z}^a})=Tr[(\Pi^{a}_{z}\otimes I_{B})\eta_{AB}]$, Bob measures coherence randomly with respect to the eigenbasis of (say) other two of the three Pauli matrices $\sigma_x$ and $\sigma_y$. As Alice's measurement in $\sigma_{k}$ basis affects the coherence of Bob's state, the coherence of the conditional state of $B$, $\eta_{B|\Pi^a_{k}}$ in the basis of $\sigma_{i}$ becomes
$C_{i}^{l_1}(\eta_{B|\Pi^a_{k}})=\sqrt{\frac{\sum_{j\neq i}\alpha_{jk_a}^2}{\gamma^2_{k_a}}},
$ where $\alpha_{ij_a}=s_{i}+(-1)^{a}t_{ji}$, $\gamma_{k_a}=1+(-1)^{a}r_k$ and $i, j, k\in \{x, y, z\}$. Note that the violation of any of the inequalities in Eq. (\ref{C_Uncer_l1}), (\ref{C_Uncer_RE}) and (\ref{C_Uncer_S}) by the conditional states of Bob implies that the single system description of coherence of the system $B$ does not exist. Thus, the criterion for achieving the non-local advantage on quantum coherence of Bob using the $l_{1}$-norm comes out to be
\begin{eqnarray}
\frac{1}{2}\sum_{\substack{i,j,a}}p(\eta_{B|\Pi_{j\neq i}^a})C_i^{l_1}(\eta_{B|\Pi_{j\neq i}^a})> \sqrt{6},
\label{l1_con}
\end{eqnarray}
 where $p(\eta_{B|\Pi_{j}^a})=\frac{\gamma_{j_a}}{2}$, $i,j\in \{x,y,z\}$ and $ a\in\{0,1\}$. This inequality forms a volume in 2-qubit state space. 
 
Let us now derive the same criterion following the relative entropy of coherence measure. We can easily 
show that the eigenvalues of the conditional state of $B$, $\eta_{B|\Pi_i^a}$ are given by 
$\lambda_{i_a}^{\pm}=\frac{1}{2}
\pm\frac{\sqrt{\sum_{j}\alpha_{ji_a}^{2}}}{2\gamma_{i_a}}$. Therefore, the relative entropy of coherence, when 
Alice measures in $\Pi_{k}^{a}$, is given by
$C^{E}_{i}(\eta_{B|\Pi_k^a})=\sum_{p=+,-}\lambda_{k_a}^{p}
\log_{2}\lambda_{k_a}^{p}-\beta_{ik_a}^{p}\log_{2}\beta_{ik_a}^{p}$,
where the diagonal element $\beta_{ij_a}^{\pm}$ of the conditional state $\eta_{B|\Pi_j^a}$, when expressed 
in the $\sigma_{i}^{th}$ basis is given by $\beta_{ij_a}^{\pm}=\frac{1}{2}\pm\frac{\alpha_{ij_a}}{2\gamma_{j_a}}$. Thus, the criterion for achieving the non-local advantage of quantum coherence becomes
(\ref{C_Uncer_RE})
\begin{eqnarray}
\frac{1}{2}\sum_{\substack{i,j,a}}p(\eta_{B|\Pi_{j\neq i}^a})C_i^{E}(\eta_{B|\Pi_{j\neq i}^a})\textgreater C^{m}_{2},
\label{RE_con}
\end{eqnarray}
 where $i,j\in \{x,y,z\}$ and $ a\in\{0,1\}$. Similarly, we obtain another inequality using the skew information as the observable measure of quantum coherence. The coherence of the conditional state 
$\eta_{B|\sigma_k^a}$ measured with respect to $\sigma_{i}$ in this case is given by
$C^{S}_{i}(\eta_{B|\Pi_k^a})=\frac{(\sum_{j\neq i}\alpha_{jk_a}^{2})(1-\sqrt{1-(2\lambda_{k_a}
^{\pm}-1)^2})}{\gamma^{2}_{k_a}(2\lambda_{k_a}^{\pm}-1)^{2}}$. Thus, from Eq. (\ref{C_Uncer_S}) we get the coherence steering inequality using the 
skew-information complementarity relation as
\begin{equation}
\frac{1}{2}\sum_{\substack{i,j,a}}p(\eta_{B|\Pi_{j\neq i}^a})C_i^{S}(\eta_{B|\Pi_{j\neq i}^a})\textgreater 2,
\label{Sk_Con}
\end{equation}
 where $i,j\in \{x,y,z\}$ and $ a\in\{0,1\}$.
 
 It is important to mention here that although the violation of the coherence complementarity relations in Eqs. (\ref{C_Uncer_l1}), (\ref{C_Uncer_RE}) and (\ref{un_skew}) implies the steerability of the quantum state and the achievability of the non-local advantage of quantum coherence, its violation highly dependent on the measurement settings \cite{supp}. Therefore, the state of Bob ($B$) can achieve the non-local advantage of quantum coherence by the help of Alice if at least one of the inequalities in Eqs. (\ref{l1_con}), (\ref{RE_con}) and (\ref{Sk_Con}) is satisfied but it is not necessary. A better choice of projective measurement bases by Alice may reveal steerability of an apparently unsteerable state with respect to the above inequalities. On the other hand, it is also necessary to show that separable states can never violate the coherence complementarity relations using the present protocol.
 
To show that no separable state can violate the coherence complementarity relations, we use the protocol stated above for arbitrary number $n$ of measurement settings. We consider a two-qubit separable state $\rho_{ab}$ as
\begin{equation}\label{sepa}
\rho_{AB}=\sum_{i}p_{i}\rho_{A}^{i}\otimes\rho_{B}^{i},
\end{equation}
with $\sum_{i}p_{i}=1$ and $p_{i}> 0$ for all $i$. Suppose, Alice performs a projective measurement in an arbitrary basis $\Pi^{a}_{n}$, where $a\in\{0, 1\}$ corresponding to two outcomes of the measurement and $n\in \mathcal{Z^{+}}$ (set of positive integers), each measurement basis associated to an integer. To compare with the coherence complementarity relations, Alice must choose $3\mathcal{Z^{+}}$ number of measurement bases, making $n$ to run upto $3k$ (say), where $k\in\mathcal{Z+}$. This provides Bob $2^{k}$ number of coherence measurement results on a particular Pauli basis. This is due to the fact that for measurements on each basis, Bob can measure coherence randomly only on two of the three mutually unbiased Pauli bases. Bob receives the state $\rho_{B|\Pi^{a}_{n}}=\frac{\sum_{i}p_{i}\langle n^{a}|\rho_{A}^{i}|n^{a}\rangle\rho_{B}^{i}}{\sum_{i}p_{i}\langle n^{a}|\rho_{A}^{i}|n^{a}\rangle}$ with probability $p(\rho_{B|\Pi^{a}_{n}})=\sum_{i}p_{i}\langle n^{a}|\rho_{A}^{i}|n^{a}\rangle$ due to the projective measurement $\Pi^{a}_{n}$ by Alice. If the proposed protocol is followed, one can show that the above state in Eq. (\ref{sepa}) can never violate the coherence complementarity relations. To show that, we start with
\begin{eqnarray}\label{proof}
&&\sum_{a=0, n=1, m=0}^{1, 3k, 1}p(\rho_{B|\Pi^{a}_{n}}) C_{n\oplus m}^{q}(\rho_{B|\Pi^{a}_{n}})\nonumber\\&&\leq
\sum_{a,n,m,i}p_{i} C_{n\oplus m}^{q}(\langle n^{a}|\rho_{A}^{i}|n^{a}\rangle\rho_{B}^{i})\nonumber\\&&\leq
\sum_{a,n,m,i}p_{i}\langle n^{a}|\rho_{A}^{i}|n^{a}\rangle C_{n\oplus m}^{q}(\rho_{B}^{i})\nonumber\\&&=\sum_{i}\sum_{n=1, m=0}^{3k, 1}p_{i} C_{n\oplus m}^{q}(\rho_{B}^{i}),
\end{eqnarray}
where we denote $n\oplus m=Mod(n+m,3)+1$ and $q\in\{l_{1}, E, S\}$, stands for various measures of coherence. In the second and the third inequalities, we used the fact that coherence and the observable measure of quantum coherence decreases under classical mixing of states. Here, we use the notation $\{C_{1}^{q}, C_{2}^{q}, C_{3}^{q}\}\equiv\{C_{x}^{q}, C_{y}^{q}, C_{z}^{q}\}$. By taking the summation over $n$ and $m$, one can show from the last line of Eq. (\ref{proof}) that
\begin{eqnarray}
&&\sum_{a, n, m}p(\rho_{B|\Pi^{a}_{n}}) C_{n\oplus m}^{q}(\rho_{B|\Pi^{a}_{n}})\nonumber\\&\leq &2^{k}\sum_{i}p_{i}\Big(C_{x}^{q}(\rho_{B}^{i})+C_{y}^{q}(\rho_{B}^{i})+C_{z}^{q}(\rho_{B}^{i})\Big)\nonumber\\ &\leq & 2^{k}\sum_{i}p_{i}\epsilon^{q}=2^{k}\epsilon^{q},
\end{eqnarray}
where $\epsilon^{q}\in\{\sqrt{6}, 2.23, 2\}$ depending on $q$. This implies that the coherence complementarity relations can never be violated by any separable state. Mathematically, for any separable state
\begin{equation}
\frac{1}{2}\sum_{a=0,n=1,m=0}^{1,3,1}p(\rho_{B|\Pi^{a}_{n}}) C_{n\oplus m}^{q}(\rho_{B|\Pi^{a}_{n}})\leq\epsilon^{q}
\end{equation}
for three measurement settings scenario ($k=1$). Violation to this inequality implies that the state is steerable and Bob can achieve non-local advantage of quantum coherence by Alice. 

Let us now illustrate the coherence steerability condition with an example, say, two qubit Werner state defined by
$\rho_w= p |\psi^{-}_{AB}\rangle\langle\psi^{-}_{AB}| +\frac{(1-p)}{4}I^{A}\otimes I^{B},$
 where $|\psi^{-}_{AB}\rangle=\frac{1}{\sqrt{2}} (|01\rangle-|10\rangle)$ and the mixing parameter $p$ is chosen from 
the range $0\leq p\leq 1$. For this state,  $\vec{r}=0$, $\vec{s}=0$, and $t_{xx}=t_{yy}=p$, $t_{zz}=-p$. The state $\rho_w$ is steerable for $p>\frac{1}{2}$, entangled for $p>\frac{1}{3}$ and Bell non-local for $p>\frac{1}{\sqrt{2}}$.

 Here, the optimal strategy for Alice to maximize the violation of coherence complementary relation by Bob's conditional state is similar to as stated earlier for the derivation of Eq. (\ref{l1_con}), (\ref{RE_con}) and (\ref{Sk_Con}). With the help of these inequalities, it is easy to show that for the Werner state, the coherence of the state of $B$ is steerable for $p>\sqrt{\frac{2}{3}}$ when one uses the $l_1$-norm as a measure of coherence, $p>0.914$ when one uses the relative entropy of coherence as a measure and $p>\frac{2\sqrt{2}}{3}$ for the choice of skew information as a measure of quantum coherence.
\begin{figure}
\resizebox{7.5cm}{6cm}{\includegraphics{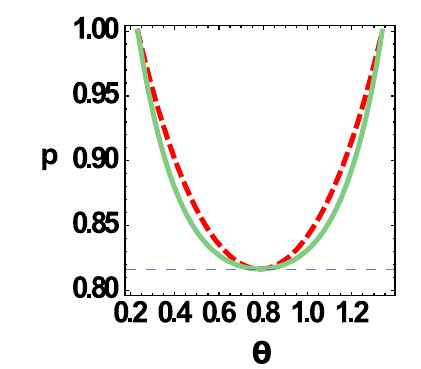}}
\caption{Filtering operation $F(\theta)=\text{diagonal}\{1/\cos(\theta),$ $1/\sin{\theta}\}$ is applied on the Werner state $\rho_{w}$. The red coloured dashed line corresponds to the situation, when $F(\theta)$ is applied on Alice and Green solid plot is when it is applied on Bob. The non-local advantage of quantum coherence is not achievable by the resulting state for the ranges of $p$ under the curves.  For example, the resulting state is steerable or the state can achieve the non-local advantage of quantum coherence for Bob for $p\geq 0.845$, when $F(\theta\approx 0.5)$ is applied on Bob. Here we assume that Alice is the steering party, and Bob is the party to be
steered. The horizontal thin dashed line denotes $p=\sqrt{\frac{2}{3}}$.}
\label{fig:2}
\end{figure} 
 
Hence, Alice controls the coherence of Bob's system for $p>\sqrt{\frac{2}{3}}$ whereas Alice controls  Bob's state for $p>\frac{1}{2}$. This difference occurs due to the presence of noise part ($\frac{I\otimes I}{4}$) in steering the state, whereas, coherence steerability criteria are never influenced by such classical noise. This raises a natural question: Is it possible to increase the range of $p$ to control the coherence of Bob's system using local filtering operations? It has been shown that filtering operations can improve the  steerablity~\cite{Brunner_Steer}. From the Fig. (\ref{fig:2}), it is clear that filtering operation on Bob can increase the range of $p$ to some extent for certain values of $\theta$, for which the resulting state can achieve the non-local advantage of quantum coherence from Alice to Bob. Moreover, any steerable Werner state can be turned into an unsteerable state by local filtering operations \cite{Brunner_Steer} (see Fig. (\ref{fig:2})).

To summarize, in this work, we use various measures of quantum coherence and derive complementarity relations~(\ref{C_Uncer_l1}),~(\ref{C_Uncer_RE})~and~
(\ref{C_Uncer_S}) between coherences of single quantum system (qubit) measured in the mutually unbiased bases. Using these complementarity relations, we derive conditions (\ref{l1_con}),
 (\ref{RE_con}) and (\ref{Sk_Con}), under which the non-local advantage of quantum coherence can be achieved for any general 
two-qubit bipartite systems. These conditions also provide a sufficient criteria for  state to be steerable. We also show that not all steerable states can achieve the non-local advantage on quantum coherence. 

Additionally, our results reveal an important connection between quantum non-locality and quantum speed limit (see supplemental material \cite{supp}). One can show that not all steerable states or for that matter, not even all states, for which non-local advantage on quantum coherence is achievable, can, in principle, achieve non-local advantage on QSL \cite{supp}. Only those states, which achieve non-local advantage on observable measure of quantum coherence or asymmetry \cite{gour1,gour2,marvian1,ahmadi} can achieve non-local QSL \cite{supp}. One important application of our results has been uncovered in the detection of Unruh effects as well \cite{chiru}.

%

{\it Note: } When this paper first appeared on arxiv, Fan \emph{et al.} presented a study on the quantum coherence of steered states~\cite{antony1} around the same time. We consider our works to be complementary. Though examining a similar topic, our approaches are very different (we consider steering from the existence of a local hidden state model rather than from the perspective of the QSE formalism).

{\it Acknowledgements:}  T.P. acknowledges financial support from ANR retour des post-doctorants NLQCC 
(ANR-12-PDOC-0022- 01). We thank Sk. Sazim for useful discussion. We also thank the anonymous referees for useful comments.

\vspace{3cm}
\onecolumngrid
{\LARGE Supplemental material:Non-Local Advantage of Quantum Coherence}
\section{I. Proof of coherence complementarity relations}To prove the coherence complementarity relation based on the $l_1-$norm, we need to prove $C^{l_1}_x C^{l_1}_y + C^{l_1}_x C^{l_1}_z+ C^{l_1}_y C^{l_1}_z \leq C_x^2 + C_y^2 +C_z^2\,\leq\, 2$. 

To prove this, let us start with the identity 
\begin{eqnarray}
(C_{x}^{l_1}-C_{y}^{l_1})^2+(C_{y}^{l_1}-C_{z}^{l_1})^2+(C_{z}^{l_1}-
C_{x}^{l_1})^2\geq  0\nonumber
\end{eqnarray}
\begin{eqnarray}
&\text{or,}&(C_x^2 + C_y^2 +C_z^2)-(C^{l_1}_x C^{l_1}_y + 
C^{l_1}_x C^{l_1}_z+ C^{l_1}_y C^{l_1}_z)\geq   
0\nonumber\\
&\text{or,}&C^{l_1}_x C^{l_1}_y + C^{l_1}_x C^{l_1}_z+ C^{l_1}_y C^{l_1}_z \leq C_x^2 + C_y^2 +C_z^2\nonumber\\&=&(n_{y}^2+n_{z}^2)+(n_{z}^2+n_{x}^2)+(n_{x}^2+n_{y}^2)
\nonumber\\&=&2
(n_{x}^2+n_{y}^2+n_{z}^2)\nonumber\\&\leq &2,
\end{eqnarray}
 where in the last line, we used the symmetry of the function, from which it is evident that the maxima is at $n_x=n_y=n_z=1/\sqrt{3}$.
 
 Now, we prove the second complementarity relation based on the relative entropy of coherence. 
 \begin{eqnarray}
\sum_{i=1}^{3}&C_{i}^{E}&=\sum_{i=1}^{3}-\frac{1+n_{i}}{2}log_{2}\frac{1+n_{i}}{2}-\frac{1-n_{i}}{2}log_{2}\frac{1-n_{i}}{2}-3\frac{1+|n|}{2}log_{2}\frac{1+|n|}{2}-3\frac{1-|n|}{2}log_{2}\frac{1-|n|}{2}\nonumber\\&=&-\sqrt{3}\frac{1+\sqrt{3}}{2}log_{2}\frac{1+\sqrt{3}}{2\sqrt{3}}-\sqrt{3}\frac{\sqrt{3}-1}{2}log_{2}\frac{\sqrt{3}-1}{2\sqrt{3}}\approx  2.23,
 \end{eqnarray}
 where in the second equality, we use the symmetry of the function to find the maxima. From the symmetry of the function, it is evident that the maxima is at $n_x=n_y=n_z=1/\sqrt{3}$.

\section{II. Measurement settings and non-local advantage of quantum coherence}
\begin{figure}
\includegraphics[scale=1.5]{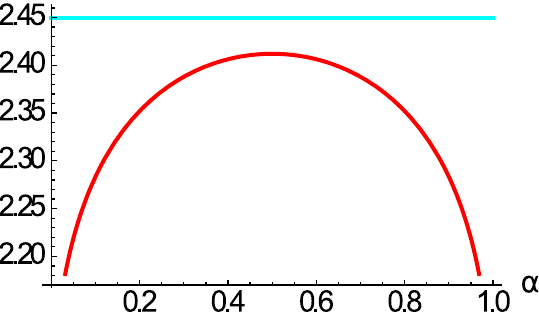}
\caption{The state $|\psi_{\alpha}\rangle_{ab}$ does not show violation of the coherence complementarity relations if Alice performs projective measurements on the Pauli bases. Although, it shows steerability if Alice chooses better measurement settings.}
\label{fig2}
\end{figure}

\begin{figure}
\includegraphics[scale=1.2]{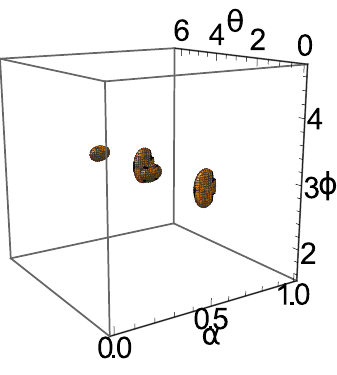}
\caption{The state $|\psi_{\alpha}\rangle_{ab}$ can be turned into a  steerable state for $\alpha$ around $\frac{1}{2}$ by performing projective measurements using arbitrary mutually unbiased bases with $\theta$ and $\phi$, which lie inside the volume. Entanglement of the state is given by linear entropy $S_{L}=\frac{(1-\alpha ) \alpha }{2 \left(\sqrt{(1-\alpha ) \alpha }+1\right)^2}$ and maximum 0.05 at $\alpha\approx 0.47$}
\label{fig3}
\end{figure}

We consider a pure entangled state $|\psi^{\alpha}_{AB}\rangle=\frac{1}{1+\sqrt{\alpha-\alpha^2}}(\sqrt{\alpha}|++\rangle+\sqrt{1-\alpha}|00\rangle)$. From the Fig. (\ref{fig2}), it is clearly visible that if Alice performs projective measurements in the Pauli bases, the state of Bob ($B$) cannot achieve the non-local advantage of quantum coherence. 
On the other hand, one can construct a set of arbitrary mutually unbiased bases as $|n_{z}^{\pm}\rangle=\cos\frac{\theta}{2}|0\rangle+e^{i\phi}\sin\frac{\theta}{2}|1\rangle$, $|n_{x}^{\pm}\rangle=\frac{|n_{z}^{+}\rangle\pm|n_{z}^{-}\rangle}{\sqrt{2}}$ and $|n_{y}^{\pm}\rangle=\frac{|n_{z}^{+}\rangle\pm i|n_{z}^{-}\rangle}{\sqrt{2}}$. If Alice performs measurements on these bases, for certain values of $\theta$ and $\phi$, as shown in fig. (\ref{fig3}), the coherence complementarity relations are violated.

From the Fig. (\ref{fig3}), it can also be seen that the states with very low amount of entanglement, cannot easily achieve non-local advantage of quantum coherence but a better measurement settings may reveal its steerability.
\section{III. Non-local effects on Quantum Speed Limits}

It is well known that quantum mechanics imposes a fundamental limit to the speed of quantum evolution, conventionally known as quantum speed limit (QSL). Suppose, a quantum system in a state $\rho_{1}$ evolves to $\rho_{2}$ under a unitary evolution operator $U$. The minimum time it takes to evolve is of fundamental interest in quantum metrology, quantum computation, quantum algorithm, quantum cryptography and quantum thermodynamics. Here, we apply our observations to understand the effects of steering kind of non-locality on the quantum speed limits (QSL) \cite{mandelstam,deba,deba1}.  Now, the development of quantum technology is at par with the advent of quantum information and computation theory. Various attempts are being made in the laboratory to implement quantum gates, which are basic building blocks of a quantum computer. Performance of a quantum computer is determined by how fast these logic gates drive the initial state to a final state. An efficient quantum gate should transform the input state into the desired state as fast as possible. This naturally presents an important issue, to address how to control and manipulate the speed of quantum evolution so as to achieve faster and efficient quantum gates. Such study may also be useful in quantum thermodynamics and other developing fields of quantum information theory. In quantum thermodynamics, this may help us to understand how to control a thermodynamic engine non-locally and use quantum correlations in our favour to construct faster yet efficient quantum engines.

 Therefore, we focus on the study of QSL in the bipartite scenario, where a part of the system is considered to be the controller of the evolution of the other part. It is well known that quantum correlations affects the evolutions of the total quantum systems. On the other hand, how a part of the system affects the evolution speed limit of the other part using the quantum correlation or non-locality is still an important unanswered issue. Here, we show that non-locality plays an important role in setting the QSL of a part of the system. In particular, we studied the effects of non-locality on quantum coherence of a part of a bi-partite system. This, in turn, clarifies the role of quantum non-locality on QSL and the intriguing connection between QSL and quantum coherence.

 Let us consider a set of three non-commuting 2-dimensional observables, $K_1$, $K_2$ and $K_3$ for qubit. Then following our result in Eq.(8), one can easily derive a complementarity relation for observable measure of quantum coherence for $K_1$, $K_2$ and $K_3$ in qubit state space. For any qubit state $\rho$, the relation takes a form
\begin{equation}\label{forspeed}
\sum_{r=1}^{3}C^{S}_{K_r}(\rho)\leq m,
\end{equation}
where $m$ is any real number and depends only on the observables. Let us explain this with an example. We consider an arbitrary observable $K=\sum_{i=x,y,z}r_{i}.\sigma_{i}$. The skew information or the observable measure of quantum coherence of an arbitrary state $\rho=\frac{1}{2}(I_{2}+\sum_{i=x,y,z}n_{i}.\sigma_{i})$, where $\sum_{i}n_{i}^2\leq 1$ with respect to the observable $K$ is given by
\begin{eqnarray}
C^{S}_{K}(\rho)=-\frac{1}{2}Tr[\sqrt{\rho},K]^2=\frac{\left(1-\sqrt{1-|n|^2}\right) |\vec{n}\times\vec{r}|^2}{|n|^2},
\end{eqnarray}
where $\vec{r}=(r_1, r_2, r_3)$ and $\vec{n}=(n_1, n_2, n_3)$. If we consider $K_{1}=\frac{1}{2}I_{2}+2\sigma_{x}$, $K_{2}=\sigma_{x}+2\sigma_{y}$ and $K_{3}=I_{2}+\sigma_{y}$, it can be easily shown that the value of the quantity $m= 10$.

We know that the evolution time bound for a state $\rho_{1}$ under a time independent Hamiltonian $H$ ($U_{H}(t)=e^{\frac{iHt}{\hbar}}$) evolving to $\rho_2=U_{H}(\tau)\rho_{1}U_{H}(\tau)^{\dagger}$ at time $\tau$ is given by (see \cite{deba})
\begin{equation}
\tau\geq T_{b}(H, \rho_{1})=\frac{\hbar}{\sqrt{2}}\frac{\cos^{-1}A(\rho_1,\rho_2)}{\sqrt{C_{H}^{S}(\rho_1)}},
\end{equation}
where $A(\rho_1,\rho_2)=\text{\rm Tr}(\sqrt{\rho_1}\sqrt{\rho_2})$ is the affinity between the initial and the final state. Let us redefine affinity as $A(\rho_1,\rho_2)=A_{H}(\rho_{1})$. 

Now, consider a bi-partite state $\rho_{AB}$ shared between Alice ($A$) and Bob ($B$). To verify Alice's control, Bob asks Alice to steer the state of system $B$ in the eigenbasis of $K_{1}$, $K_{2}$ or $K_{3}$. Bob measures the QSL of the conditional state of $B$, $\rho_{A|\Pi_{K_1}}$ by evolving the state under the unitary evolution governed by  $K_{2}$ or $K_{3}$ in case of the claim by Alice that she controls her state in the eigenbasis of $K_{1}$ and so on. Thus, using Eq. (\ref{forspeed}), one can easily show that
\begin{eqnarray}
\frac{1}{2}\sum_{\substack{r,s,a}}p(\rho_{A|\Pi_{K_{s\neq r}}^{a}})\left(\frac{\cos^{-1}A_{K_{r}}(\rho_{A|\Pi_{K_{s\neq r}}^{a}})}{T_{b}(K_{r}, \rho_{A|\Pi_{K_{s\neq r}}^a})}\right)^2\leq \frac{2m}{\hbar^2},
\end{eqnarray}
where $r, s\in \{1, 2, 3\}$ and $a\in \{0,1\}$.

Violation of this inequality, for a set of observables $K_{1}$, $K_{2}$ and $K_{3}$ implies non-local advantage on QSL achievable for the state. In particular, let us consider Werner state $\rho_w= p |\psi^{-}_{AB}\rangle\langle\psi^{-}_{AB}| +\frac{(1-p)}{4}I^{A}\otimes I^{B}$.  For the given set of observables, $K_{1}$, $K_{2}$ and $K_{3}$ if one follows the protocol, one can easily show that the state never achieves the non-local advantage on quantum speed limit although the state achieves the non-local advantage of quantum coherence. On the other hand, if one used $K_{1}=\sigma_{x}$, $K_{2}=\sigma_{y}$ and $K_{3}=\sigma_{z}$, non-local advantage of QSL could have been achieved. 

Now, consider the maximally mixed two qubit state $\frac{I_{2}^{B}\otimes I_{2}^{B}}{4}$ and the maximally entangled pure two qubit state $|\psi\rangle=\frac{1}{\sqrt{2}}(|10\rangle+|01\rangle)$. For both of the examples, the state of Bob is nothing but the maximally mixed qubit state $\frac{I_{2}}{2}$. Still, for the second state, we can achieve the non-local advantage of QSL or coherence on Bob by LOCC on Alice's system but this is not possible for the maximally mixed two-qubit state.

\end{document}